\documentclass[aps,twocolumn,floatfix]{revtex4-1}

\usepackage[dvipdfmx]{graphics,graphicx}

\usepackage{color}
\usepackage{wrapfig}
\usepackage{enumerate}
\usepackage{amssymb,amsmath}
\usepackage{bm}
\usepackage{here}

\usepackage{xspace}
\newcommand{\Hz}{\mathcal{H}^{z}_{\rm eff}\xspace}
\newcommand{\Hx}{\mathcal{H}^{x}_{\rm eff}\xspace}

\twocolumngrid
\begin{document}
\unitlength = 1mm
\title{
Chiral spin liquids at finite temperature in a three-dimensional Kitaev model
}

\author{
  Yasuyuki~Kato$^1$, 
  Yoshitomo~Kamiya$^2$,
  Joji~Nasu$^3$,
  and Yukitoshi~Motome$^1$}
\affiliation{
  $^1$Department of Applied Physics, University of Tokyo, Hongo, 7-3-1, Bunkyo, Tokyo 113-8656, Japan\\
  $^2$Condensed Matter Theory Laboratory, RIKEN, Wako, Saitama 351-0198, Japan\\
  $^3$Department of Physics, Tokyo Institute of Technology, Meguro, Tokyo 152-8551, Japan
}

\date{\today}

\begin{abstract}
  Chiral spin liquids (CSLs) in three dimensions and thermal
   phase transitions to paramagnet are studied 
   by unbiased Monte Carlo simulations. 
   For an extension of the Kitaev model to a three-dimensional tricoordinate network dubbed the hypernonagon lattice,
  we derive low-energy effective models in two different anisotropic limits. 
  We show that the effective interactions between the emergent $Z_2$ degrees of freedom called fluxes are unfrustrated in one limit, while highly frustrated in the other.
  In both cases, we find a first-order phase transition to the CSL, where 
  both time-reversal and parity symmetries are spontaneously broken.
  In the frustrated case, however, the CSL state is highly exotic --- 
  the flux configuration is subextensively degenerate while showing a directional order with broken $C_3$ rotational symmetry.
  Our results provide two contrasting archetypes of CSLs in three dimensions, both of which allow approximation-free simulation for the thermodynamics. 
\end{abstract}
\maketitle

\section{Introduction}
The quantum spin liquid (QSL) is a long-standing subject, investigated for more than 40 years~\cite{anderson1973}. 
Recently, it attracted renewed attention not merely within basic science~\cite{balents2010,lacroix2011} but also
due to its relevance to quantum computations~\cite{kitaev2003,nayak2008}.
The chiral spin liquid (CSL), which is the subject of this paper, belongs to a special subgroup of QSLs with 
spontaneous breaking of time-reversal ($\mathcal{T}$) symmetry.
It has been a key concept in condensed matter physics, e.g., the fractional quantum Hall effect~\cite{laughlin1990},
high-$T_c$ superconductivity~\cite{anderson1987,wen1989},
frustrated quantum Heisenberg models~\cite{kalmeyer1987,wen1989,messio2012,bauer2014,gong2014},
and braiding of anyonic elementary excitations in QSLs~\cite{kitaev2006,yao2007}.

Recently, a new trend in the study of CSLs has been created by exactly soluble models in the ground state~\cite{kitaev2006,yao2007,Schroeter2007,hermanns2017,dusuel2008}. 
This trend was initiated by an intriguing suggestion by Kitaev~\cite{kitaev2006}: 
On a tricoordinate network with odd-site loops, one can construct a model that realizes an exact CSL ground state.
Indeed, a quantum spin model on a decorated honeycomb network, which has triangles in the lattice structure, was exactly shown to have the 
CSL ground state~\cite{yao2007}; 
the CSL can be either topologically trivial or nontrivial depending on the exchange couplings,
accommodating Abelian or non-Abelian anyonic excitations, respectively~\cite{yao2007}.
The nature of the finite-temperature ($T$) phase transitions to these topologically different CSLs was also elucidated
by using a quantum Monte Carlo simulation~\cite{nasu2015}.

Compared to these studies of CSLs in two dimensions (2D), much less is known in three dimensions (3D). 
Nevertheless, 3D CSLs are intriguing 
because of exotic excitations specific to 3D, such as anyonic loop excitations of emergent fluxes~\cite{si2008}
and Weyl semimetallic excitations of Majorana fermions~\cite{obrien2016}.
These possibilities make the study of 3D CSLs at finite $T$ even more interesting, including transitions breaking 
parity ($\mathcal{P}$) symmetry as well as $\mathcal{T}$ symmetry.
While loop like excitations in the 3D Kitaev models and other realizations of 3D $Z_2$ QSL are known to trigger a thermal second-order phase transition~\cite{Castelnovo2008,Mandal2014,nasu2014,nasu2014b,kamiya2015},
rather than a crossover in the case of 2D $Z_2$ QSL~\cite{nasu2015b}, the transitions to 3D CSLs remain elusive thus far.

In this paper, 
we present unbiased numerical results for 3D CSLs and thermal phase transitions to paramagnet. 
We consider an extension of the Kitaev model~\cite{kitaev2006} defined on a three-dimensional tricoordinate network labeled by (9,3)a 
in the classification of Wells~\cite{wells1977,obrien2016}, 
which we call the hypernonagon lattice because the elementary loop consists of nine bonds.
We derive the low-energy effective models for two distinct anisotropic limits, which are described  
by interacting $Z_2$ fluxes.
We find that the effective model in one limit has no frustration while that in the other limit is highly frustrated.
Using Monte Carlo (MC) simulations, we show that both models undergo a first-order phase transition from high-$T$ paramagnet to a low-$T$ CSL, where both $\mathcal{T}$ and $\mathcal{P}$ symmetries are spontaneously broken.
Interestingly, neither of the two cases yields a uniform flux configuration in the low-$T$ CSL states unlike in the 2D case~\cite{yao2007}.
Of particular interest is the frustrated case: The CSL has subextensive accidental degeneracy in the flux configuration,
while exhibiting a directional order with breaking of $C_3$ rotational symmetry 
in addition to $\mathcal{T}$ and $\mathcal{P}$ symmetries.

This paper is organized as follows.
In Sec.~II, we introduce the extended Kitaev model on the hypernonagon lattice 
and derive the low-energy effective Hamiltonians 
in two distinct anisotropic limits.
We also describe the MC method for investigating the thermodynamic behavior of the two low-energy models. 
In Sec.~III, we present the MC results of 
thermodynamic behaviors of the two models as well as the analysis of the ground state properties.
Finally, Sec.~IV is devoted to the summary.

\section{Models and method}
\begin{figure}[!t] 
  \centering
  \includegraphics[width=\columnwidth,trim = 40 0 20 0,clip]{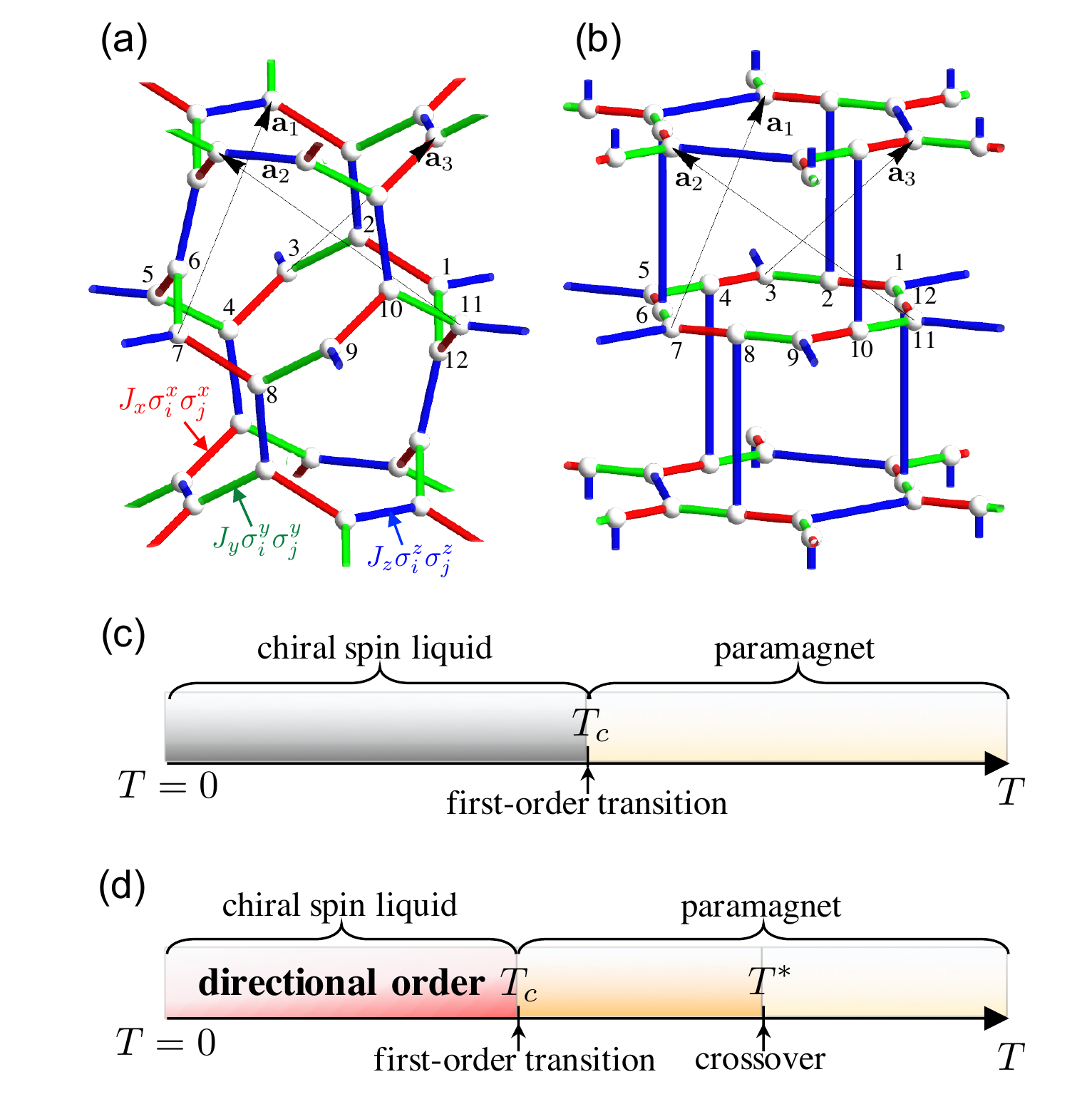}
  \caption{
    (a) Hypernonagon-lattice Kitaev model and (b) the alternative visualization~\cite{obrien2016}.
     ${\bf a}_1$, ${\bf a}_2$, and ${\bf a}_3$ are the primitive vectors. 
    The spheres represent the $S=1/2$ spins.
    The sublattice indices for 12 spins in a unit cell are shown.
    Schematic finite-$T$ phase diagrams in (c) the large $J_z$ limit and (d) the large $J_x$ limit.
  }
  \label{lattice}
\end{figure}

\subsection{Kitaev model on the hypernonagon lattice}
We consider a straightforward extension of the Kitaev model~\cite{kitaev2006} on the hypernonagon lattice shown in Fig.~1(a).
The most noteworthy characteristics of this lattice distinct from many other 3D tricoordinate lattices is that it has odd-site loops.
Such odd-site loops accommodate emergent $Z_2$ fluxes that are odd under both $\mathcal{T}$ and $\mathcal{P}$ operations, and hence, the ground state of the system can be a CSL~\cite{kitaev2006}.
The Hamiltonian of the hypernonagon Kitaev model 
is given by
\begin{eqnarray}
  \mathcal{H}  =
  \sum_{\mu =\{ x,y,z \} } \mathcal{H}_\mu,~~~
  \mathcal{H}_\mu =
  -J_\mu \sum_{\langle i,j \rangle_\mu} \sigma^\mu_i \sigma^\mu_j,
\label{eq:model}
\end{eqnarray}
where $\sigma_i^\mu$ ($\mu=x,y,z$) is a Pauli matrix ($\sigma^z_i = \pm 1$) at site $i$
and the sum $\sum_{\langle i,j \rangle_\mu}$ runs over all the nearest neighbors connected by $\mu$ bonds shown in Fig.~\ref{lattice}(a)
[see also Fig.~\ref{lattice}(b)].
The number of elementary nine-site loops
is eight per unit cell,
and the centers of these loops can be combined into a ``cube,'' as shown in Fig.~\ref{bplattice}(a).
Each loop center is shared by two different types of cubes,
denoted as ``B'' and ``R'', as shown in Fig.~\ref{bplattice}(a). 
These corner-sharing cubes form a 3D version of the checkerboard lattice as shown in Fig.~2(b).

For each nine-site loop, one can define the $Z_2$ flux operator 
\begin{eqnarray}
	W_p = -i \prod_{ \langle i,j \rangle_\mu \in p} \sigma^\mu_i \sigma^\mu_j,
\end{eqnarray}
where the product is taken for all the bonds in the loop $p$ in a clockwise manner
viewed from the center of each B cube~[Fig.~\ref{bplattice}(a)]. 
$W_p$ is a conserved quantity which is odd under both $\mathcal{T}$ and $\mathcal{P}$ operations with the eigenvalues 
$\pm 1$ (called $\pm \pi/2$ flux~\cite{obrien2016}).
Similar to other 3D cases~\cite{obrien2016,Mandal2014}, there are 
local constraints on $W_p$ corresponding to the operator identities for Pauli matrices:
The product of eight $W_p$ is always 
unity in each B and R cube.
Thus, the eigenstates of the model in Eq.~(\ref{eq:model}) are divided into the sectors with different configurations of $W_p$, and hence, the ground state can be,
in principle, obtained by comparing the eigenenergies.
According to the variational calculation, however, the hypernonagon model has complexity: Low-energy sectors are nearly degenerate when $J_x \sim J_y \sim J_z$~\cite{obrien2016}.

\subsection{Low-energy effective Hamiltonians in two anisotropic limits}
\begin{figure}[!htb]
  \centering
  \includegraphics[width=\columnwidth]{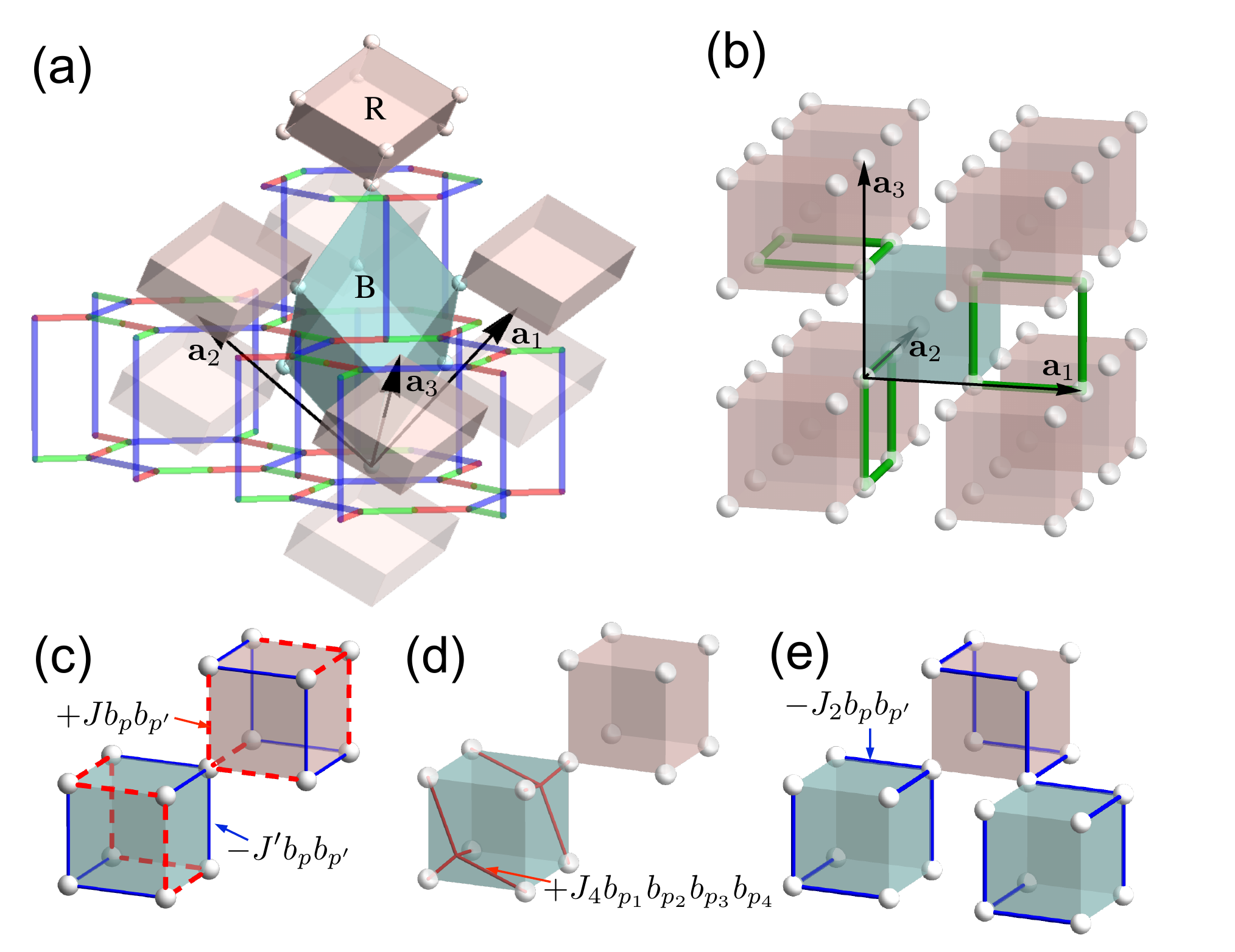}
  \caption{
      Relation between the hypernonagon lattice and the lattice of $b_p$ in the effective models in Eqs.~\eqref{eq:heffz} and \eqref{eq:heffx}.
    (a) A  distorted cubic lattice where the spheres represent $b_p$.
    (b) A 3D checkerboard lattice topologically equivalent to (a).
      The green squares represent examples of the four-site loops of $b_p$ for MC update.
      The interactions between $b_p$ for 
      (c) the large $J_z$ limit and (d,e) the large $J_x$ limit.
      }
  \label{bplattice}
\end{figure}

We derive low-energy effective Hamiltonians of Eq.~(\ref{eq:model}) in two different anisotropic limits: 
the large $J_z$ limit ($J_z \gg J_x, J_y$) and the large $J_x$ limit ($J_x \gg J_y, J_z$)
\footnote{The large $J_y$ limit is equivalent to the large $J_x$ limit by symmetry.}. 
Following the derivation of the toric code for the honeycomb Kitaev model~\cite{kitaev2006},
we perform the perturbation expansion in terms of $\mathcal{H} - \mathcal{H}_{\mu}$ 
for the unperturbed Hamiltonian $\mathcal{H}_{\mu}$.
The effective Hamiltonians can be written in terms of $Z_2$ variables describing the flux states for each loop.
By the expansion up to the eighth order, 
we obtain the following effective Hamiltonians, $\mathcal{H}^{z}_{\rm eff}$ and $\mathcal{H}^{x}_{\rm eff}$, for the large $J_z$ and $J_x$ limits, respectively:
\begin{eqnarray}
	\mathcal{H}_{\rm eff}^z &=& J \sum_{\langle p,p' \rangle} b_p b_{p'} - J' \sum_{( p,p' )} b_p b_{p'}\nonumber\\
	&&-J^{(4)} \sum_{( p_1, p_2, p_3, p_4 )}  b_{p_1}b_{p_2}b_{p_3}b_{p_4},
	 \label{eq:heffz}\\	
	\mathcal{H}_{\rm eff}^x &=& J_4 \sum_{\langle p_1,p_2,p_3,p_4 \rangle} b_{p_1} b_{p_2}b_{p_3}b_{p_4} - J_2 \sum_{( p,p' )} b_p b_{p'},\label{eq:heffx}
\end{eqnarray}
with
\begin{eqnarray}
&&J=\frac{33}{2048} \frac{J_x^4 J_y^4}{|J_z^7|},~~
J'= \frac{9}{33} J,~~
J^{(4)}=\frac{1}{2} J,
\\
&&J_2=\frac{9}{2048} \frac{J_y^4 J_z^4}{|J_x^7|},~~
J_4= \frac{63}{512} \frac{J_y^6}{|J_x^5|}.
\end{eqnarray}
$J$, $J'$, and $J_2$ are obtained by
the eighth-order perturbation, while $J_4$ is the sixth-order one
\footnote{For $J_4$, we only consider the leading contribution.}. 
See Appendix A for details of the derivation.
Here, $b_p$ is a $Z_2$ variable defined as 
\begin{eqnarray}
  b_p =  \mathcal{P}_\mu W_p \mathcal{P}^{\;}_\mu = \pm 1,
\end{eqnarray}
where $\mathcal{P}_\mu$ is the projection to the ground state manifold of $\mathcal{H}_{\mu}$.
The models include no odd-order term in $b_p$, precluded by $\mathcal{T}$ and $\mathcal{P}$ symmetries. 
The sums $\sum_{\langle p,p' \rangle}$ and $\sum_{(p,p')}$ run over the specific bonds indicated by solid blue and dashed red lines,
respectively, in Figs.~\ref{bplattice}(c) and \ref{bplattice}(e),
while 
$\sum_{(p_1,p_2,p_3,p_4)}$ in Eq.~(\ref{eq:heffz}) and
$\sum_{\langle p_1,p_2,p_3,p_4\rangle}$ in Eq.~(\ref{eq:heffx})
 run over 
 all the faces of B and R cubes where $p_1$--$p_4$ indicate the corners of each square face 
 and
 ``clusters'' comprising four $b_p$ as shown in Fig.~\ref{bplattice}(d), respectively.
Similar to $W_p$, $b_p$ obeys the local constraints,
i.e., the product of eight $b_p$ in each cube must be unity. 
In addition, there are two global constraints, similar to the hyperhoneycomb case~\cite{Mandal2014,kato2017}.

\subsection{Monte Carlo method}
As both $\Hz$ and $\Hx$ are given in terms of the static $Z_2$ variables $b_p$, 
their thermodynamic properties can be investigated by classical MC simulations, similar to Ref.~\onlinecite{nasu2014}.
To satisfy the local and global constraints discussed above, a pair of four-site loops of $b_p$ 
must be flipped simultaneously in a single update in the MC simulation~\cite{kato2017}.
Examples of the four-site loops are shown in Fig.~\ref{bplattice}(b).
We adopt the annealing technique unless otherwise noted.
The observables and statistical errors are evaluated from 24--384 independent sets of $10^5$--$10^7$ MC samples.

\section{Results}
In this section, we present the results of analysis of the low-energy effective Hamiltonians on their ground states and thermodynamic behaviors for the both large $J_z$ and large $J_x$ limits.
\subsection{Large $J_z$ limit}
\subsubsection{Ground state}
Let us first discuss the effective model in the large $J_z$ limit, 
$\Hz$ in Eq.~\eqref{eq:heffz}. 
As $J$ and $J^{(4)}$ are relatively larger than $J'$ in the model in Eq.~(\ref{eq:heffz}), the lowest-energy flux configuration is a simple staggered one shown in the inset of Fig.~\ref{physQz}. 
The ground state energy per $b_p$ is simply computed as
\begin{eqnarray}
 \varepsilon^z_{\rm GS} =  
 -\frac{ 3}{2} ( J - J' + J^{(4)}) 
 = -\frac{81}{44} J.
 \label{eq:gse_Jz}
\end{eqnarray}

\subsubsection{Monte Carlo simulation at finite temperature}
\begin{figure}[!htb]
  \centering
  \includegraphics[width=\columnwidth,trim= 40 300 440 0, clip]{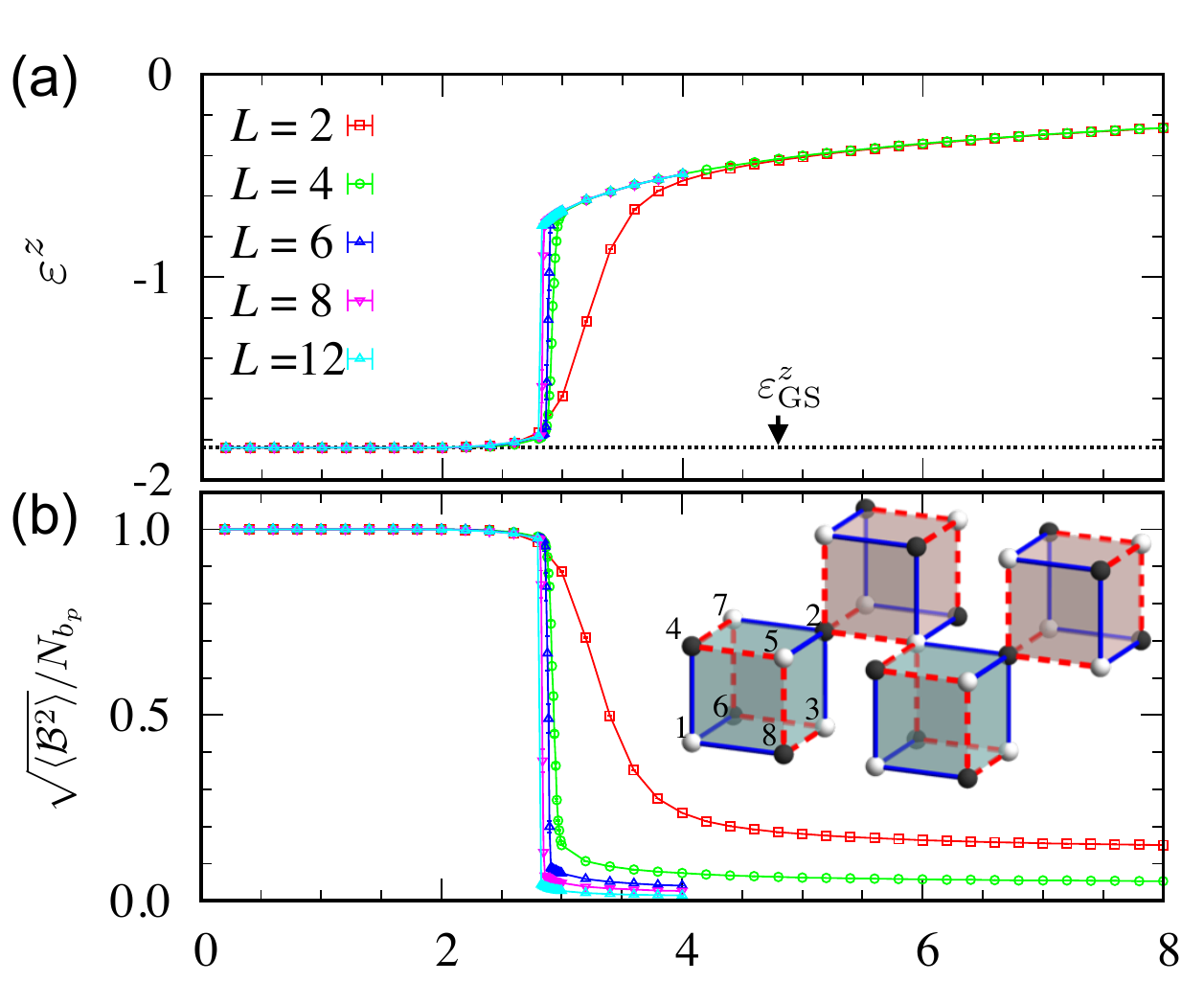}
  \caption{
     Temperature dependence of (a) the energy density $\varepsilon^z$
    and (b) the order parameter $\sqrt{ \langle \mathcal{B}^2 \rangle }/N_{b_p}$
    for the large-$J_z$ effective model $\mathcal{H}^z_{\rm eff}$ in Eq.~\eqref{eq:heffz}.
    We set $J = 1$ ($J' = 9/33$, $J^{(4)}=1/2$). 
    $\varepsilon^z_{\rm GS}$ indicates the ground state energy in Eq.~(\ref{eq:gse_Jz}) 
    for the $b_p$ configuration in the inset of (b);
    the black (white) spheres represent $b_p = -1(+1)$
    [or $b_p=+1(-1)$], and the numbers denote the sublattices. 
  }
  \label{physQz}
\end{figure}

Figure~\ref{physQz} shows the MC results for the large
$J_z$ model. 
We find that the system undergoes a phase transition at $T_c \simeq 3 J$  with a discontinuous jump in the energy density  
$\varepsilon^z \equiv \langle \Hz \rangle /N_{b_p}$ ($N_{b_p} = 8 L^3$ is the number of $b_p$ sites and
$L$ is the linear dimension of the $b_p$ lattice in Fig.~\ref{bplattice}).
Below $T_c$, the staggered magnetization, defined as
$\sqrt{ \langle \mathcal{B}^2 \rangle } / N_{b_p}$ where $\mathcal{B} = \sum_p (-1)^p b_p$,
becomes nonzero with a jump from $0$ to $\approx 1$ in the thermodynamic limit.
These observations indicate that the system undergoes a strong first-order transition from the paramagnetic phase to the CSL phase~[Fig.~\ref{lattice}(c)].

Regarding this discontinuous behavior, both the constraints on $b_p$ and the peculiar symmetry of $b_p$ defined on odd-site loops must play a central role. 
Since a similar strong first-order phase transition to a CSL
is seen in the MC simulations with $J^{(4)}=0$, 
the four-body interaction is not the origin of the strong discontinuity.
In addition, without the constraints and the $J^{(4)}$ term, 
$\mathcal{H}^z_{\rm eff}$ is merely an unfrustated Ising model, 
which undergoes a continuous transition.
Similarly,  when the system is composed of even-site loops,
the leading term in perturbation theory is linear in the flux variable, 
which can also be mapped onto an unfrustrated Ising model by a duality transformation~\cite{nasu2014,kamiya2015}.

\subsection{Large $J_x$ limit}
\subsubsection{Ground state}
\label{sec:Jx-gs}
\begin{figure}
  \centering
  \includegraphics[trim = 0 0 0 0, clip,width=\columnwidth]{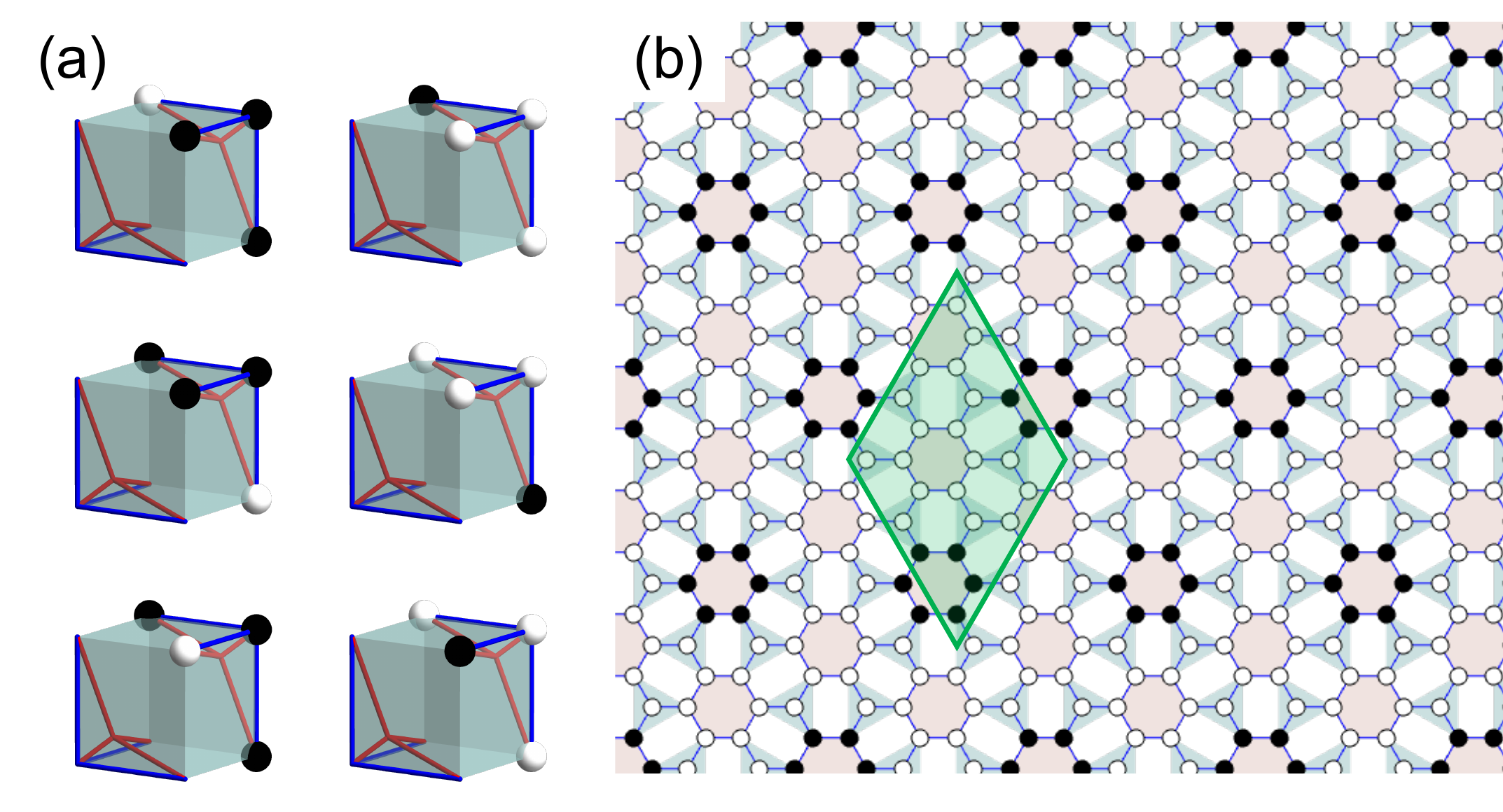} 
  \caption{
  	(a) Six configurations in a four-flux cluster minimizing its intracluster interaction energy ($J_4$ and $J_2$ interaction).
  	(b) An example of the $b_p$ configurations in the ground state manifold 
  	on a (111) hexagon-triangular plane in the large $J_x$ limit.
	The rhombic region indicated by the green solid line is the unit cell.
	 \label{figS:gsconfig}
  }
\end{figure}

Next we discuss the effective model in the large $J_x$ limit, $\Hx$ in Eq.~\eqref{eq:heffx}. 
In contrast to the large $J_z$ model $\Hz$, the model suffers from frustration, and the ground state manifold exhibits substantial degeneracy for $J_4 \gg J_2$ (note that $J_4$ is in the lower-order perturbation than $J_2$).
First of all, the four-body interactions in the $J_4$ terms must be optimized:
$b_{p_1} b_{p_2} b_{p_3} b_{p_4} = -1$ in every four-flux cluster $\langle{p_1, p_2, p_3, p_4}\rangle$ shown in Fig.~\ref{bplattice}(d).
Any of the resulting configurations corresponds to a $\pi$-flux state,
in contrast to the 0-flux state in the large $J_z$ limit~\cite{obrien2016}.
In addition to this condition, the ground state manifold satisfies the following three energetics (i)--(iii).
First, (i) $J_2$ favors six configurations in each four-flux cluster shown 
in Fig.~\ref{figS:gsconfig}(a); here we note that the $\pi$-flux states cannot optimize all the $J_2$ terms simultaneously.
Also note that the local constraint associated with a given B cube is fulfilled 
for any combination of the six states for a pair of four-$b_p$ clusters per B cube.
Meanwhile, (ii) the six-site network of $J_2$ within each R cube [Fig.~\ref{bplattice}(e)] favors six $b_p$
on the buckled hexagon ($h$) to be either all $+1$ or all $-1$.
Finally, the energetics (ii) also implies that (iii) the two remaining $b_p$ on each R cube, i.e., not on the hexagon $h$, 
[for example $p_a$ and $p_b$ in the inset of Fig.~\ref{gsconfig}(a)] 
must take the same value because of the local constraint of R cube.

On the basis of the consideration above, we obtain the ground state energy of the effective model in Eq.~\eqref{eq:heffx} as follows.
The largest contribution to the ground state energy is $-J_4$ per 4 $b_p$, i.e., $-J_4/4$ per $b_p$, from the $J_4$ terms.
To count the energy contribution from the $J_2$ term, 
let us consider an example of the $b_p$ configurations which satisfy the energetics (i)--(iii).
For this purpose, it is convenient to view the 3D checkerboard lattice from the [111] direction,
and to extract a layer of $b_p$ connected by the $J_2$ bonds;
the system can be regarded as a stacking of ``hexagon-triangular'' layers, as shown in Fig.~\ref{figS:gsconfig}(b). 
The black and white circles in Fig.~\ref{figS:gsconfig}(b) 
exemplifies a ground state configuration in a (111) hexagon-triangular plane, whose unit cell is relatively small 
(including 24 $b_p$ and 36 $J_2$ bonds, as shown by the green rhombus). 
The two-body interactions in the $J_2$ term are satisfied on the 30 bonds, while unsatisfied on the 6 bonds.
Thus, the energy contribution from the $J_2$ term is $(6-30)J_2 = -24 J_2$ per unit cell, i.e.,   $-J_2$ per $b_p$.
The ground state $b_p$ configurations must satisfy also the energetics (iii) arising from the local constraint.
This is readily satisfied by stacking the optimized $b_p$ configurations like in Fig.~\ref{figS:gsconfig} in a proper manner. 
Thus, we find the ground state energy per $b_p$ in the large $J_x$ limit as
\begin{eqnarray}
\varepsilon^x_{\rm GS} = -J_2 - \frac{J_4}{4}.
\label{eq:gse_Jx}
\end{eqnarray}

\subsubsection{Monte Carlo simulation at finite temperature}
\begin{figure}[!htb]
  \centering
  \includegraphics[width=\columnwidth]{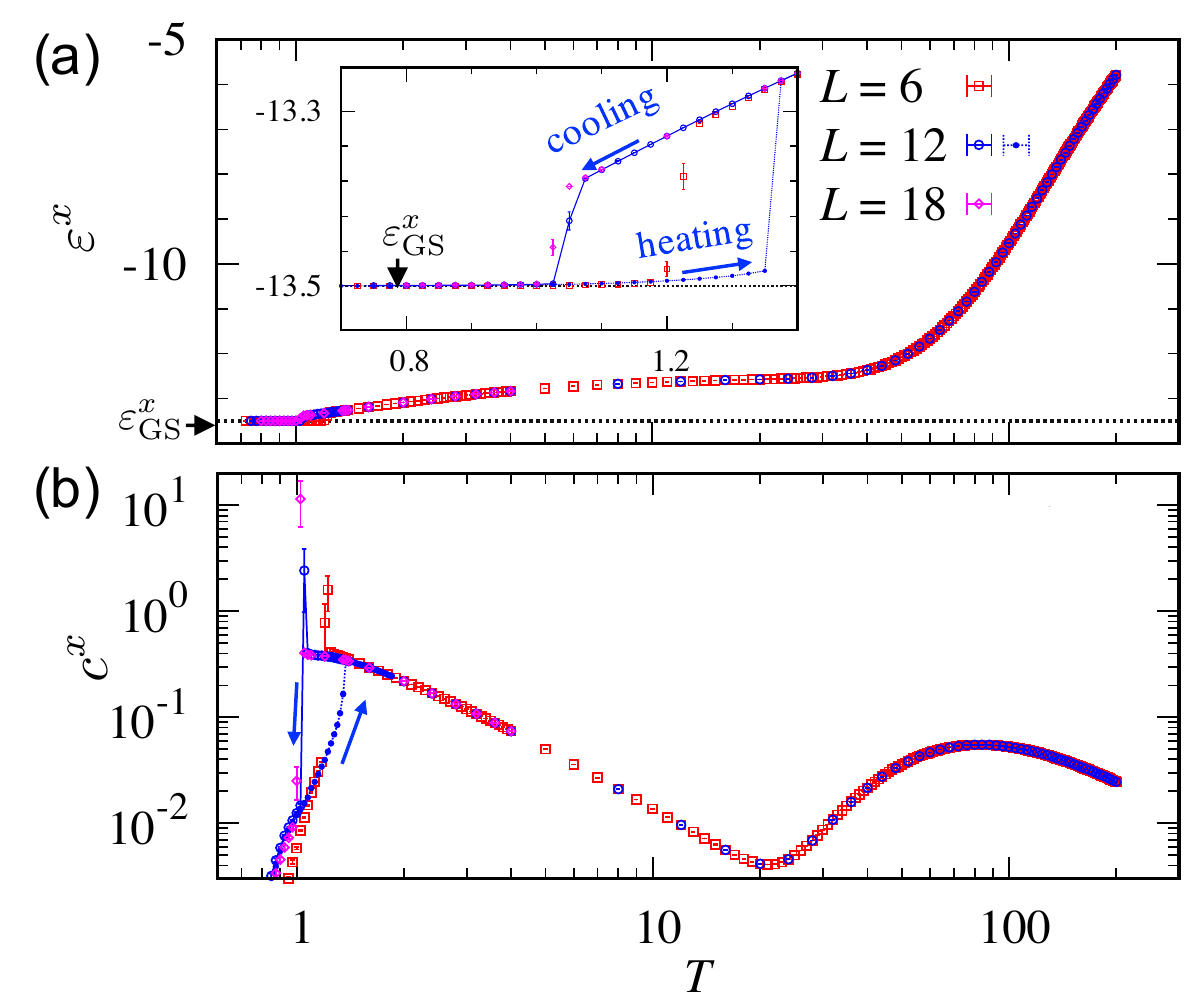}
  \caption{
    Temperature dependence of (a) the energy density $\varepsilon^x$ and  
    (b) the specific heat $c^x$
      for the large-$J_x$ effective model $\mathcal{H}^x_{\rm eff}$ in Eq.~(\ref{eq:heffx}).
    We set $J_2 = 1$ and $J_4 = 50$.
      $\varepsilon^x_{\rm GS}$ indicates the ground state energy in Eq.~(\ref{eq:gse_Jx}). 
    The inset of (a) shows an enlarged view of the main panel.
    The data of cooling and heating processes are for demonstrating the hysteresis (the heating process is shown only for $L=12$).
  }
  \label{physQx}
\end{figure}
Figure~\ref{physQx} shows the MC results for $\Hx$, where we set $J_2/J_4 = 0.02$ considering that $J_2$ is higher order in perturbation theory than $J_4$.
While decreasing $T$, there are two successive drops in $\varepsilon^x = \langle{\Hx}\rangle /N_{b_p}$ at $T^* \sim J_4$
and $T_c \sim J_2$. 
Correspondingly, the specific heat
$c^x = \partial \varepsilon^x /\partial T$ exhibits a broad peak at $T^*$ and a sharp peak at $T_c$.
$T^*$ is a crossover temperature, below which
configurations with $b_{p_1} b_{p_2} b_{p_3} b_{p_4} = 1$ is exponentially suppressed in every four-flux cluster $\langle{p_1, p_2, p_3, p_4}\rangle$ shown in Fig.~\ref{bplattice}(d).
Upon further decreasing $T$, the three local energetics (i)--(iii) discussed above emerge.

In fact, the singularity at $T_c$ signals a transition to a CSL state in which the above (i)--(iii) are all satisfied in the $T = 0$ limit.
As evidenced by the hystereses in $\varepsilon^x$ and $c^x$ in Fig.~\ref{physQx}, this transition is also of first order.

\begin{figure*}[!htb]
  \centering
  \includegraphics[width=\textwidth,trim= 0 100 0 0,clip]{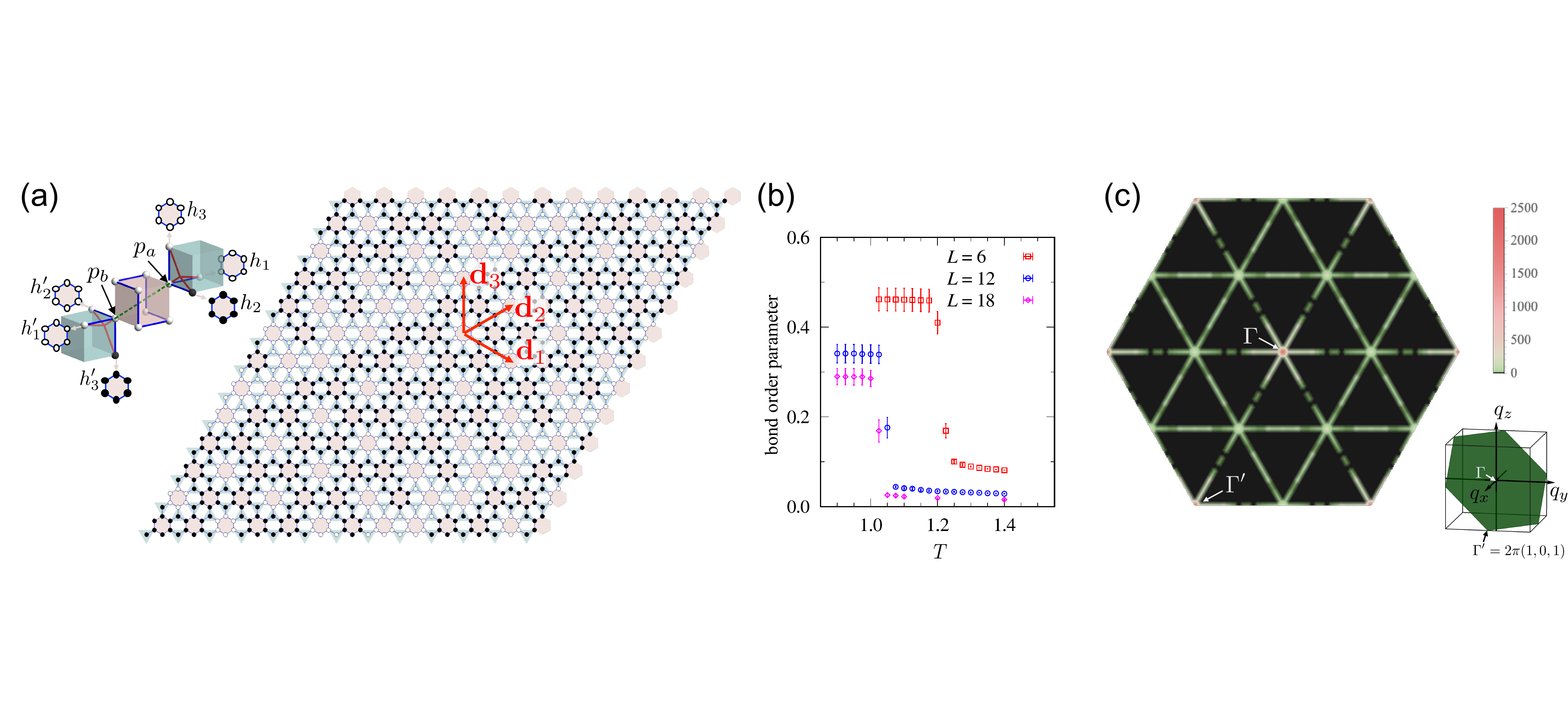}
  \caption{
    (a) MC snapshot of the $b_p$ configurations below $T_c$ on a 
    hexagon-triangular layer corresponding to a (111) slice of the 3D checkerboard lattice in Fig.~\ref{bplattice}.
    The black (white) circles represent $b_p = -1(+1)$.
    The inset illustrates an example of a favored configuration for a pair of four-$b_p$ $J_4$-clusters on the second adjacent (111) hexagon-triangular layers at low $T$.
    (b) Temperature dependence of the bond order parameter with respect to $B_h$ defined in Eq.~\eqref{eq:bop}.
    (c) Structure factor for $b_p$ on the plane of $q_x+q_y+q_z=0$ (see the inset). 
    The data are obtained below $T_c$ by averaging over $\sim 300$ MC samples for $L=18$.
  }
  \label{gsconfig}
\end{figure*}

As discussed in Sec.~\ref{sec:Jx-gs}, the energetics (i)-(iii) cannot select out an ordered configuration of $b_p$, and leave subextensive degeneracy.
In the MC simulation below $T_c$, 
we also find the 
subextensive degeneracy in the $b_p$ configurations, 
along with spontaneous breaking of a point-group symmetry below $T_c$. 
To explain this, 
we show a MC snapshot on a (111) hexagon-triangular plane in Fig.~\ref{gsconfig}(a).
Here, the hexagons are the $J_2$ networks in each R cube, in most of which
$B_h \equiv \sum_{p \in h} b_p /6 = \pm 1$ below $T_c$ because of the energetics (ii). 
In a given hexagon-triangular layer, three buckled hexagons, say $h_1$, $h_2$, and $h_3$ forming a triangle, are interconnected by a four-flux $J_4$-cluster in a B cube~[see the inset of Fig.~\ref{gsconfig}(a)]. 
Because of the frustrated energetics (i), the ground state has $B_{h_1}B_{h_2}B_{h_3} = -1$ for any triangle, resembling the situation in the triangular-lattice Ising model~\cite{Wannier1950}. 
However, unlike this classic problem, the flux configurations generated by MC simulation appear
to break $C_3$ rotational symmetry; an example is shown in Fig.~\ref{gsconfig}(a). 
We confirm the $C_3$ breaking by measuring the bond order parameter with respect to $B_h$
defined as follows.
At first, we consider a direction specific correlator of $B_h$ as 
\begin{equation}
r_\nu \equiv \frac{1}{N_h} \sum_{h} B_h B_{h + {\bf d}_\nu },
\end{equation}
where ${\bf d}_\nu$ ($\nu=1,2,3$) are the inplane vectors shown in Fig.~\ref{gsconfig}(a),
the sum $\sum_{h}$ runs over all the hexagons $h$ in every second (111) layers (hexagon-triangular layers) connected by the effective interaction,
and $N_h$ is the number of the hexagons.
Then, we define the bond order parameter as
\begin{eqnarray}
\frac{1}{2} \Big(3 r_{\max} - \sum_{\nu=1}^3 \langle r_\nu \rangle\Big),\label{eq:bop}
\end{eqnarray}
where
$r_{\max} = \langle \max [ r_1, r_2, r_3 ]\rangle$.
As plotted in Fig.~\ref{gsconfig}(b), the bond order parameter becomes finite
below $T_c$, which is an indication of the directional order 
selecting one of three directions ${\bf d}_\nu$ shown in Fig.~\ref{gsconfig}(a). 
Likewise, in the structure factor
 \begin{eqnarray}
 S({\bf q}) \equiv  \frac{1}{N_{b_p}} \langle b_{\bf q} b_{-{\bf q}}\rangle,
 \end{eqnarray}
 with $b_{\bf q} = \sum_p b_p e^{-i {\bf q}\cdot {\bf r}_p}$ 
 (${\bf r}_p$ is the position vector for the site $p$), 
 we find diffusive lines in $S({\bf q})$ consistent with the directional order. 

This ``locking transition'' is suggested to be induced by the interlayer coupling, similar to the Ising model on the stacked triangular layers~\cite{Blankschtein1984,Coppersmith1985,Matsubara1987,Moessner2000,Isakov2003,Jiang2005,Lin2014}. 
Coming back to the consideration of the energetics (iii), the two $b_p$ in a R cube not included in a buckled hexagon
[$b_{p_a}$ and $b_{p_b}$ in the inset of Fig.~\ref{gsconfig}(a)] also belong to four-flux $J_4$-clusters that are on second adjacent layers.
As each of them combines three buckled hexagons (say, $h_1$--$h_3$ and $h'_1$--$h'_3$) on each $(111)$ honeycomb-triangular layer, the energetics (iii) implies an effective interlayer coupling favoring $(B_{h_1} + B_{h_2} + B_{h_3})(B_{h'_1} + B_{h'_2} + B_{h'_3}) = +1$.
This is expected to play an important role in the locking transition; in fact,
$S({\bf q} = 0)$ is divergent below $T_c$. 
This is also an indication of breaking of $\mathcal{T}$ and $\mathcal{P}$ symmetries in the low-$T$ CSL.

Thus, in the large $J_x$ limit, the system exhibits a first-order transition similar to the large $J_z$ limit, but the low-$T$ CSL state is not completely ordered while it has the directional order 
with the uniform component of $b_p$~[Fig.~\ref{lattice}(d)].
The CSL phase is highly unusual ---
it is not ordered in the double meaning: The original spins $\sigma_i$ in Eq.~(\ref{eq:model}) are disordered, and in addition, the emergent $Z_2$ fluxes $b_p$ are not completely ordered.
However, it is characterized by a directional order with broken $C_3$ rotational symmetry.
The peculiar nature may yield more exotic elementary excitations than ever studied in 3D CSLs.

\section{Summary}
In summary, we discovered two distinct 3D CSLs,
both of which allow unbiased simulations for the thermodynamics. 
We showed that one of them suffers from severe frustration in interacting $Z_2$ fluxes. 
By unbiased Monte Carlo simulations, we found that both 
CSLs undergo a first-order phase transition to paramagnet.
Remarkably, the frustrated CSL retains degeneracy while showing a directional order.
Our discovery of two interesting cases will stimulate further studies of 3D CSLs. 
Nature of elementary excitations will be an intriguing future issue, especially for the 
exotic directionally-ordered CSL.

\begin{acknowledgments}
The authors thank S. Trebst and M. Hermanns for stimulating discussion in the early stage of this study.
This work was supported by JSPS Grant 
No.~26800199, No.~JP15K13533, No.~JP16K17747, No.~JP16H02206, and No.~JP16H00987.
Numerical calculations were conducted on the supercomputer system in ISSP, The University of Tokyo.
\end{acknowledgments}

\appendix
\section{Derivation of the low-energy effective Hamiltonian}
\begin{figure}
  \centering
  \includegraphics[trim = 0 0 0 0, clip,width=\columnwidth]{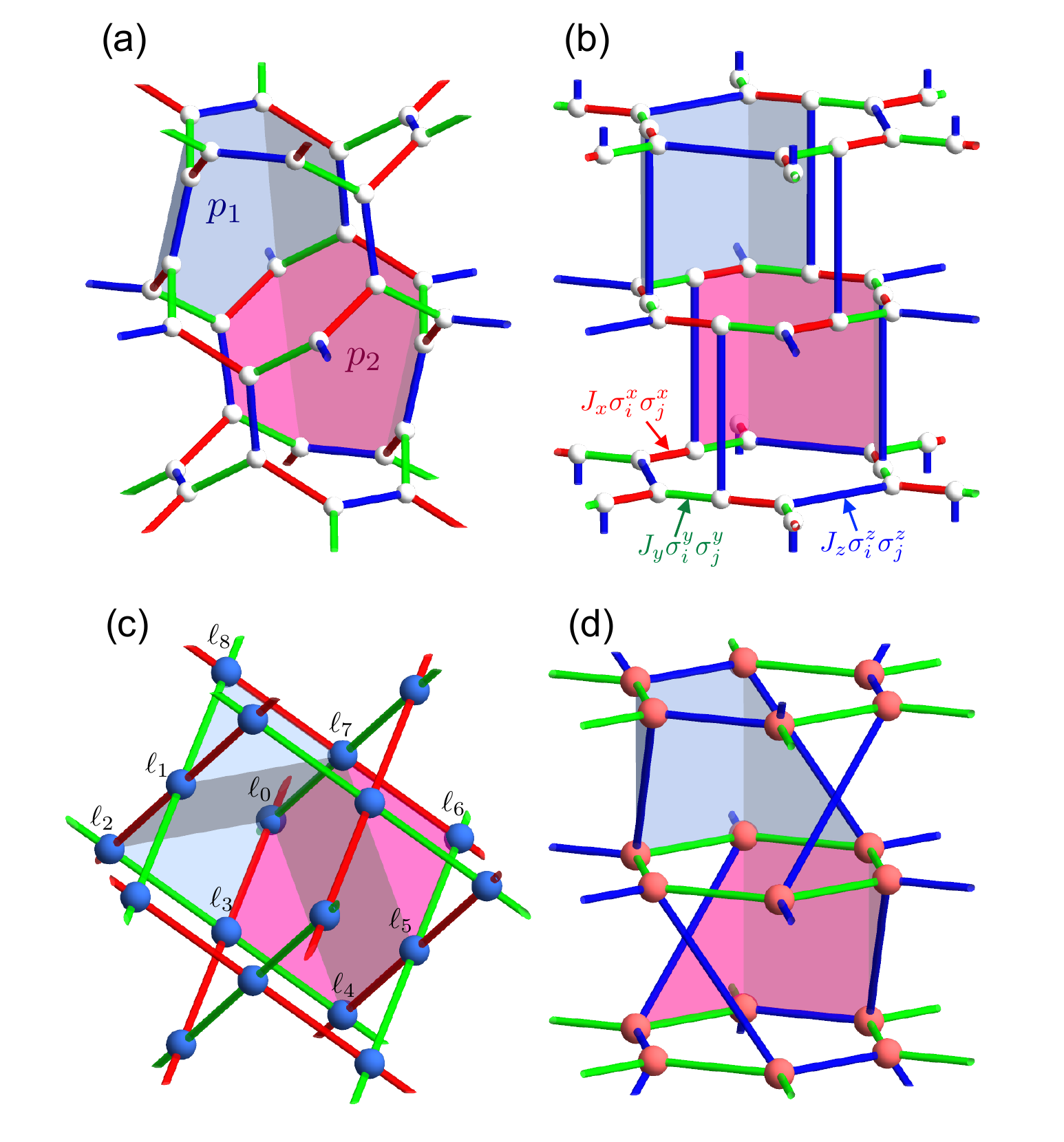} 
  \caption{
    (a)(b) The hypernonagon lattice reproduced from Figs.~\ref{lattice}(a) and \ref{lattice}(b).
    The blue and red plaquettes represent a neighboring pair 
    of the nine-site loops ($p_1$ and $p_2$), on which 
    the local conserved quantities $W_p$ are defined.
    (c) A layered Lieb lattice 
    [(d) a layered honeycomb lattice] 
    obtained by contracting all the $z$ bonds ($x$ bonds) for the large $J_z$ ($J_x$) limit.
    The blue and red plaquettes represent $b_{p_1}$ and $b_{p_2}$, respectively.
    The reddish and bluish spheres represent the sites where the $\tau$ degree of freedom in the low-energy effective models is defined.
	\label{figS:lattice}
  }
\end{figure}
In this Appendix, 
we show how to derive the low-energy effective Hamiltonians in Eqs.~\eqref{eq:heffz} and \eqref{eq:heffx}. 
We derive the effective Hamiltonians from the Kitaev model on the hypernonagon lattice 
in Eq.~\eqref{eq:model} 
for the large $J_z$ and large $J_x$ limits, by following the way to derive the toric code in the 
anisotropic limit of the original Kitaev model on a honeycomb lattice~\cite{kitaev2006}.
In the large $J_\mu$ limit ($\mu=z$ or $x$), we regard $\mathcal{H}_0 = \mathcal{H}_{\mu}$ 
and the rest $\mathcal{H}_1 = \mathcal{H} - \mathcal{H}_0$ 
as an unperturbed Hamiltonian and a perturbation, respectively.
The unperturbed states for $\mathcal{H}_0$ are composed of the independent dimers on the $\mu$ bonds. 
Each dimer is described by a new spin 1/2 degree of freedom ${\bm \tau}$, and 
the ground state for $\mathcal{H}_0$ is given by a direct product of the states
$\left| \tau_{ij}^z = \pm 1 \right\rangle = \left| \sigma_i^\mu = {\rm sgn}(J_\mu) \sigma_j^\mu = \pm 1 \right\rangle$ for all the $\mu$ bonds ($\langle i, j\rangle_\mu$).
When we define ${\bm \tau}$ at the center of each $\mu$ bond, the lattice structure for the ${\bm \tau}$ degree of freedom looks like Figs.~\ref{figS:lattice}(c) and \ref{figS:lattice}(d) for the large $J_z$ and $J_x$ limits, respectively: The blue $z$ (red $x$) bonds in Figs.~\ref{figS:lattice}(a) and \ref{figS:lattice}(b) are replaced by the blue (red) sites. 
The former is regarded as a layered Lieb lattice, while the latter a layered honeycomb lattice.

When we introduce $\mathcal{H}_1$ as the perturbation, the $n$th-order contribution to the low-energy effective Hamiltonian is given by
\begin{equation}
\mathcal{H}^{(n)}_{\mu} =
\mathcal{P}^{\;}_\mu \left[
\left( \mathcal{H}_1 
\mathcal{S} 
\right)^{n-1} \mathcal{H}_1
\right]\mathcal{P}^\dag_\mu,
\label{eq:Hn}
\end{equation}
where $\mathcal{P}_\mu$ is the projection to the low-energy subspace spanned by
the direct product of the states 
$\left| { \tau_{ij}^{z} = \pm 1 }\right\rangle$;
\begin{equation}
\mathcal{S} 
\equiv 
\frac{1- \mathcal{P}^\dag_\mu \mathcal{P}^{\;}_\mu }{E_0 - \mathcal{H}_0},
\end{equation}
where $E_0$ is the ground state energy of $\mathcal{H}_{0}$. 
The effective Hamiltonians in Eqs.~\eqref{eq:heffz} and \eqref{eq:heffx} 
are obtained by using Eq.~(\ref{eq:Hn}) up to the eighth-order perturbation.
We note that Eq.~(\ref{eq:Hn}) is not generic but valid for sufficiently low orders of the expansion. 
For example, the generic form for the fourth-order contributions is obtained as~\cite{takahashi1977}
\begin{eqnarray}
&& \mathcal{P}^{\;}_\mu \left[
\left( \mathcal{H}_1 \mathcal{S} \right)^3 \mathcal{H}_1
\right]\mathcal{P}^\dag_\mu\nonumber \\
&& -\frac{1}{2} \left[
\mathcal{P}^{\;}_\mu 
\mathcal{H}_1 \mathcal{S}^2 \mathcal{H}_1
\mathcal{P}^\dag_\mu
\mathcal{P}^{\;}_\mu 
\mathcal{H}_1 \mathcal{S} \mathcal{H}_1
\mathcal{P}^\dag_\mu \right. \nonumber \\
&& \quad \left.+
\mathcal{P}^{\;}_\mu 
\mathcal{H}_1 \mathcal{S} \mathcal{H}_1
\mathcal{P}^\dag_\mu
\mathcal{P}^{\;}_\mu 
\mathcal{H}_1 \mathcal{S}^2 \mathcal{H}_1
\mathcal{P}^\dag_\mu
\right]. 
\label{eq:per}
\end{eqnarray}
The second term in Eq.~(\ref{eq:per}) is omitted in Eq.~(\ref{eq:Hn}).
Since the $n$th-order perturbation lower than or equal to the eighth-order in the large $J_z$ case 
(the sixth order in the large $J_x$ case)
leads to only constants, we neglect 
the contributions from the second term in Eq.~(\ref{eq:per}) in the following calculations.

The derivation of the effective models in Eqs.~\eqref{eq:heffz} and \eqref{eq:heffx}
is lengthy but straightforward. 
For instance, let us consider the two-body $J$ term in Eq.~\eqref{eq:heffz}. 
It is derived from the eight-site loop $\ell_{1}$-$\ell_2$-$\cdots$-$\ell_{8}$ in Fig.~\ref{figS:lattice}(c). 
The eight-site loop is made of two neighboring six-site elementary loops $p_1$ (blue plaquette $\ell_8$-$\ell_7$-$\ell_0$-$\ell_3$-$\ell_2$-$\ell_1$) and $p_2$ (red plaquette $\ell_4$-$\ell_3$-$\ell_0$-$\ell_7$-$\ell_6$-$\ell_5$), as shown in
Fig.~\ref{figS:lattice}(c).
By the perturbation on this eight-site loop 
[eighth-order perturbation in $\mathcal{H}_1$ by using Eq.~(\ref{eq:Hn})],
we obtain
\begin{equation}
-J
\tau_{\ell_1}^y
\tau_{\ell_2}^z
\tau_{\ell_3}^x
\tau_{\ell_4}^y
\tau_{\ell_5}^y
\tau_{\ell_6}^z
\tau_{\ell_7}^x
\tau_{\ell_8}^y. 
\label{eq:Jterm}
\end{equation}
The blue and red plaquettes are originally derived from those in Figs.~\ref{figS:lattice}(a) and \ref{figS:lattice}(b), on which the $Z_2$ conserved quantities $W_p$ are defined. 
Hence, we can rewrite Eq.~(\ref{eq:Jterm}) by using the $Z_2$ variables $b_p$ 
which are defined as the projection of $W_p$: 
$b_p = \mathcal{P}_\mu W_p \mathcal{P}^{\dag}_\mu$.
For the blue and red plaquettes, $b_p$ are given as
\begin{eqnarray}
b_{p_1} &=& 
\tau_{\ell_8}^y
\tau_{\ell_7}^y
\tau_{\ell_0}^z
\tau_{\ell_3}^z
\tau_{\ell_2}^z
\tau_{\ell_1}^y, 
\\
b_{p_2}&=&-
\tau_{\ell_4}^y
\tau_{\ell_3}^y
\tau_{\ell_0}^z
\tau_{\ell_7}^z
\tau_{\ell_6}^z
\tau_{\ell_5}^y. 
\end{eqnarray}
Thus, the eighth-order perturbation term in Eq.~\eqref{eq:Jterm} is rewritten into the two-body interaction $J b_{p_1} b_{p_2}$.
The other interaction terms in Eqs.~\eqref{eq:heffz} and \eqref{eq:heffx} 
can be derived in a similar manner.
We note that the combination of $b_{p_1}$ and $b_{p_2}$ in the large $J_x$ limit 
corresponds to a ten-site loop as shown in Fig.~\ref{figS:lattice}(d), 
and thus there is no interaction between $p_1$ and $p_2$ in Eq.~\eqref{eq:heffx} 
within the eighth-order perturbation.

\bibliographystyle{apsrev4-1}

\bibliography{alhk}

\end{document}